\documentclass{jps-cp}
\usepackage{txfonts,bm}

\title{
Total Hadronic Cross Sections via the Holographic Pomeron Exchange
}

\author{Akira \textsc{Watanabe}$^{1,2}$}

\inst{$^{1}$Institute of High Energy Physics and Theoretical Physics Center for Science Facilities, Chinese Academy of Sciences, Beijing 100049, People's Republic of China \\
$^{2}$University of Chinese Academy of Sciences, Beijing 100049, People's Republic of China}

\email{akira@ihep.ac.cn}

\recdate{January 15, 2019}

\abst{The analysis on the hadron-hadron scattering at high energies in the framework of holographic QCD is presented.
Combining the Pomeron exchange kernel and the hadron-Pomeron couplings which are obtained from the bottom-up AdS/QCD models, we calculate the total cross sections.
Our calculation for the nucleon-nucleon scattering is consistent with the experimental data including the recent ones measured by the TOTEM collaboration at the LHC.
Within the framework, one can consider other processes involving other hadrons.
As examples to show this feature of the model, the resulting total cross sections of the pion-nucleon and pion-pion scattering are also presented.
}

\kword{QCD, gauge/string duality, AdS/CFT correspondence, LHC}

\begin{document}

\maketitle

%%%%%%%%%%%%%%%%%%%%%%%%%%%%%%
\section{Introduction}
%%%%%%%%%%%%%%%%%%%%%%%%%%%%%%
Via various high energy scattering phenomena we can investigate the internal structure of hadrons.
The recent developments in the experimental technique have provided us with the valuable opportunities to deepen our knowledge about the partonic structure of the nucleon, which is usually described by the parton distribution functions (PDFs).
The PDFs are expressed with the two kinematic variables, the Bjorken scaling variable $x$ and the energy scale $Q^2$.
It is impossible to directly measure the PDFs by experiments, because the experimental measurements are always cross sections which are basically mixtures made of the quark and the gluon contributions.
Since the perturbative technique of QCD is applicable in some limited kinematic region, where $x$ is moderate and $Q^2$ is large, the theoretical analysis with the data have developed our understandings about the nucleon structure.
However, the analysis based on the QCD itself is extremely difficult at small $x$, because the contributions from so many gluons with tiny momenta become dominant in this region.
Hence, an effective approach based on a model plays an important role in revealing the nature of the nucleon in this regime.
In general, the Pomeron exchange, which can be interpreted as the multi-gluon exchange, is often considered in the small $x$ region, and it is known that various data are well explained by this picture.

One of the most important candidates of the effective approach for the QCD analysis in the nonperturbative kinematic region is the holographic QCD, which is constructed based on the AdS/CFT correspondence~\cite{Maldacena:1997re,Gubser:1998bc,Witten:1998qj}.
The Pomeron exchange can be realized as the Reggeized graviton exchange in holographic QCD, and a lot of studies have been done for the analysis on the high energy scattering with this description.
In particular, the deep inelastic scattering (DIS) in the small $x$ region has been intensively studied~\cite{Brower:2010wf,Watanabe:2012uc,Watanabe:2013spa,Watanabe:2015mia}.
It was found that the data taken at HERA can be well described by assuming the holographic Pomeron exchange, which strongly supports further applications.

In this brief report, based on Ref.~\cite{Watanabe:2018owy} we present our analysis on the high energy hadron-hadron scattering in the framework of holographic QCD.
We combine the Pomeron exchange kernel proposed by Brower, Polchinski, Strassler, and Tan (BPST)~\cite{Brower:2006ea} and gravitational form factors obtained from the bottom-up AdS/QCD models, and calculate the total cross sections, focusing on the high energy region in which the involved gluonic strong interaction can be well approximated by the Pomeron exchange.
We show that our calculation for the nucleon-nucleon scattering agrees with the experimental data including the recent ones measured by the TOTEM collaboration at the LHC.
In our model setup, we can consider other processes involving other hadrons, because the participating hadron is characterized only by the form factor.
As examples, the results for the pion-nucleon and pion-pion cases are also presented.

%%%%%%%%%%%%%%%%%%%%%%%%%%%%%%
\section{Model setup}
%%%%%%%%%%%%%%%%%%%%%%%%%%%%%%
Following the optical theorem, the total cross section can be expressed with the scattering amplitude in the forward limit as
\begin{equation}
\sigma_{tot} (s) = \frac{1}{s} \mathrm{Im} \mathcal{A} (s, t=0),
\end{equation}
where the $s$ and $t$ are the Mandelstam variables.
Using the BPST Pomeron exchange kernel $\chi$, the scattering amplitude of the two-body scattering process, $1 + 2 \to 3 + 4$, is written in the five-dimensional AdS space as
\begin{equation}
{\mathcal A} (s, t) = 2 i s \int d^2 b e^{i \bm{k_\perp } \cdot \bm{b}} \int dzdz' P_{13}(z) P_{24}(z') \left[ 1 - e^{i \chi (s, \bm{b}, z, z')} \right],
\label{eq:amp}
\end{equation}
where $z$ and $z'$ are the fifth coordinates, and $\bm{b}$ is the two-dimensional impact parameter.
$P_{13} (z)$ and $P_{24} (z')$ are functions describing the density distributions of the involved hadrons, and they are normalized because the considered hadrons in this study are normalizable modes:
\begin{equation}
\int dz P_{13} (z) = \int dz' P_{24} (z') = 1.
\end{equation}
Picking up the leading contribution from the eikonal representation in Eq.~\eqref{eq:amp} and assuming the conformal limit, the impact parameter integration can be performed analytically.
The total cross section is then given by
\begin{equation}
\sigma_{tot} (s) = \frac{g_0^2 \rho^{3/2}}{8 \sqrt{\pi}} \int dzdz' P_{13} (z) P_{24} (z') (zz') \mathrm{Im} [\chi (s,z,z')],
\label{eq:tcs}
\end{equation}
where $g_0$ and $\rho$ are the adjustable parameters which control the magnitude and the energy dependence of the total cross section, respectively.

The analytic form of the imaginary part of the BPST kernel can be obtained in the conformal limit.
However, it was found in the preceding studies on DIS at small $x$ that some modification is needed to reproduce the experimental data especially in the small $Q^2$ region.
When the hard scale $Q^2$ increases, the QCD coupling decreases, which implies that the difference between QCD and CFT may decrease in the large $Q^2$ region.
On the other hand, at small $Q^2$ the confinement effect of QCD becomes more important and the soft physics shall be taken into account.
The corresponding hard scale in the nucleon-nucleon scattering considered in this study is the nucleon mass, $m_N \sim 1$~GeV, which is small enough.
Hence, we adopt the modified BPST kernel instead of the conformal one.
The modified kernel consists of the conformal kernel and an added term which mimics the confinement effect.
In practice, the added term suppresses the cross section in the small $Q^2$ region, and this is required to well reproduce the data.

To perform the numerical evaluations, we need to specify the density distributions, $P_{13} (z)$ and $P_{24} (z')$, in Eq.~\eqref{eq:tcs}.
They are extracted from the hadron-Pomeron(graviton)-hadron three-point functions.
In Refs.~\cite{Abidin:2008hn,Abidin:2008ku,Abidin:2009hr}, the authors calculated the gravitational form factors of hadrons by using the bottom-up AdS/QCD models~\cite{Erlich:2005qh,Hong:2006ta}.
In this study, we utilize their results from the hard-wall models, in which the AdS geometry is sharply cut off in the infrared region to introduce the QCD scale, to specify the density distributions of the nucleon~\cite{Abidin:2009hr} and the pion~\cite{Abidin:2008hn}.
Only the $z(z')$ dependence of the integrand in their resulting expressions for the gravitational form factors is required, because we focus on the forward limit.

%%%%%%%%%%%%%%%%%%%%%%%%%%%%%%
\section{Numerical results}
%%%%%%%%%%%%%%%%%%%%%%%%%%%%%%
In our model setup, there are only two adjustable parameters, $g_0$ and $\rho$ in Eq.~\eqref{eq:tcs}, which are to be determined with the experimental data.
The density distributions in Eq.~\eqref{eq:tcs} originally have a few parameters, but they can be uniquely fixed by the corresponding hadron properties such as hadron masses.
For the numerical fitting, we focus on the energy range, $10^2 < \sqrt{s} < 10^5$~GeV, and take into account the recent data taken by the TOTEM collaboration at the LHC~\cite{Antchev:2013gaa,Antchev:2013iaa,Antchev:2013paa,Antchev:2015zza,Antchev:2016vpy,Nemes:2017gut,Antchev:2017dia} and other $pp$ and $\bar{p}p$ data summarized by the Particle Data Group in 2010~\cite{Nakamura:2010zzi}.

\begin{figure}[tb]
\centering
\includegraphics[width=0.72\textwidth]{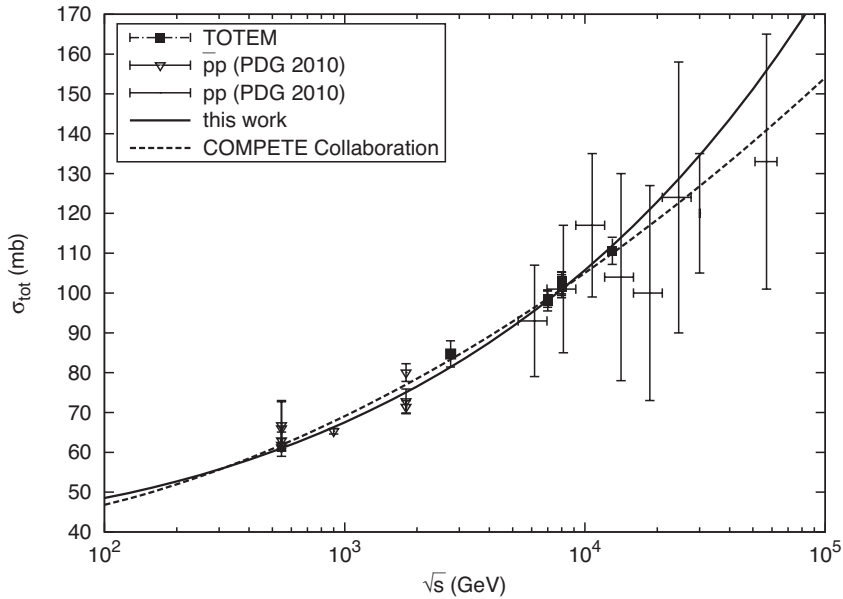}
\caption{
The nucleon-nucleon total cross section as a function of $\sqrt{s}$.
Our calculation is represented by the solid curve, and the dashed curve denotes the empirical fit obtained by the COMPETE collaboration~\cite{Cudell:2002xe}.
The experimental data~\cite{Antchev:2013gaa,Antchev:2013iaa,Antchev:2013paa,Antchev:2015zza,Antchev:2016vpy,Nemes:2017gut,Antchev:2017dia,Nakamura:2010zzi} are depicted with error bars.
}
\label{fig}
\end{figure}

We show in Fig.~\ref{fig}
the resulting nucleon-nucleon total cross section, compared to the experimental data and the empirical fit obtained by the COMPETE collaboration~\cite{Cudell:2002xe}.
It is seen that our calculation agrees with the data in the whole considered region, and is consistent with the empirical fit at $\sqrt{s} < 10$~TeV, although the substantial difference between the two curves is observed in the higher energy region.

Next, we present our results for the pion-nucleon and pion-pion scattering.
Once we determine the two adjustable parameters through the analysis of the nucleon-nucleon scattering, we can predict the total cross sections involving other hadrons without any additional parameter by replacing the density distributions.
Since the energy dependence of the obtained cross sections is totally governed by the Pomeron exchange kernel in the present model setup, we can see the differences only in the magnitudes.
Hence, it is useful to consider the ratios of the resulting total cross sections.
Our results are as follows:
\begin{equation}
\frac{\sigma_{tot}^{\pi N}}{\sigma_{tot}^{N N}} = 0.63, \ \
\frac{\sigma_{tot}^{\pi \pi}}{\sigma_{tot}^{N N}} = 0.45.
\end{equation}
There is no available data for comparison in the considered high energy region.
However, the authors of Ref.~\cite{Donnachie:1992ny} studied various hadron-hadron total cross sections, considering the soft Pomeron exchange, and it is possible to extract the ratios from their results.
The extracted value is $\sigma_{tot}^{\pi N} / \sigma_{tot}^{N N} = 0.63$ which is consistent with our prediction.

%%%%%%%%%%%%%%%%%%%%%%%%%%%%%%
\section{Summary and discussion}
%%%%%%%%%%%%%%%%%%%%%%%%%%%%%%
In this study, we have investigated the hadron-hadron total cross sections at high energies in the framework of holographic QCD.
In our model setup, the complicated gluonic strong interaction is described by the Pomeron exchange in the five-dimensional AdS space utilizing the BPST kernel, and the hadron-Pomeron gravitational couplings are obtained from the bottom-up AdS/QCD models of the considered hadrons.
The comparison between our calculation and the experimental data has been explicitly demonstrated.

The resulting nucleon-nucleon total cross section agrees with the data including the recently measured TOTEM ones in the whole considered kinematic range, $10^2 < \sqrt{s} < 10^5$~GeV.
Our result is also consistent with the empirical fit obtained by the COMPETE collaboration at $\sqrt{s} < 10$~TeV, but a substantial enhancement is observed for our calculation in the higher energy region.
To pin down the cross section in this regime, more data are certainly needed.

We have also presented our predictions for the pion-nucleon and pion-pion scattering in the same energy region.
The resulting total cross section ratio for the pion-nucleon case, $\sigma_{tot}^{\pi N} / \sigma_{tot}^{N N}$, agrees with that extracted from the results of the analysis based on the soft Pomeron exchange.
It is interesting that this observation may be consistent with the nontrivial relationship between the nucleon and the pion $F_2$ structure functions at small $x$ studied in Ref.~\cite{Watanabe:2012uc}.

The presented framework is basically applicable to other high energy scattering processes, in which the involved strong interaction can be approximated by the Pomeron exchange.
Further investigations are needed.

\end{document}